# Distinct transient structural rearrangement of ionized water revealed by XFEL X-ray pump X-ray probe experiment

Michal Stransky[1,2], Thomas J. Lane[3,4]*, Alexander Gorel[5], Sébastien Boutet[6], Ilme Schlichting[5]*, Adrian P. Mancuso[1,7,8], Zoltan Jurek[3,4], and Beata Ziaja[3,2]*

[1]European XFEL, Holzkoppel 4, 22869, Schenefeld, Germany

[2]Institute of Nuclear Physics, Polish Academy of Sciences, Radzikowskiego 152, 31-342, Krakow, Poland

[3]Center for Free-Electron Laser Science CFEL, Deutsches Elektronen-Synchrotron DESY, Notkestr. 85, 22607, Hamburg, Germany

[4]The Hamburg Centre for Ultrafast Imaging, Luruper Chaussee 149, 22761, Hamburg, Germany

[5]Max Planck Institute for Medical Research, Jahnstr. 29, 69120 Heidelberg, Germany

[6]SLAC National Accelerator Laboratory, Menlo Park, CA 94025, USA

[7]Department of Chemistry and Physics, La Trobe Institute for Molecular Science, La Trobe University, Melbourne, Victoria 3086, Australia

[8]Present address: Diamond Light Source, Harwell Science and Innovation Campus, Didcot OX11 0DE, UK

*thomas.lane@desy.de, *Ilme.Schlichting@mpimf-heidelberg.mpg.de, *Beata.Ziaja-Motyka@cfel.de

**Abstract**

Using X-ray free electron laser (XFEL) radiation to conduct an X-ray pump X-ray probe experiment, we studied strongly ionized water as part of our ongoing work on radiation damage. After irradiance with a pump pulse with a nominal fluence of $\sim 5\times 10^5$ J/cm$^2$ , we observed for pump-probe delays of 75 fs and longer an unexpected structural rearrangement, exhibiting a characteristic length scale of ~9 Å. Simulations suggest that the experiment probes a superposition of ionized water in two distinct regimes. In the first, fluences expected at the X-ray focus create nearly completely ionized water, which as a result becomes effectively transparent to the probe. In the second regime, out of focus pump radiation produces $O^{1+}$ and $O^{2+}$ ions, which rearrange due to Coulombic repulsion over 10s of fs. Importantly, structural changes in the low fluence regime have implications for the design of two-pulse X-ray experiments that aim to study unperturbed liquid samples. Our simulations account for two key observations in the experimental data: the decrease in ambient water signal and an increase in low-angle X-ray scattering. They cannot, however, account for the experimentally observed 9 Å feature. A satisfactory account of this feature presents a new challenge for theory.

# Introduction

Theoretical approaches are well-established for condensed matter, gases, and plasmas. Recently, however, it has become clear that at densities and temperatures falling in between these established regimes, matter can adopt unique properties that theory cannot fully account for[1,2]. Often referred to as warm dense matter (WDM), these states are characterized by densities between $10^{-2}$ and $10^4$ g/cm$^3$ and temperatures on the order of $10^3$ - $10^7$ Kelvin (0.1-1000 eV), as found in brown dwarf stars, the cores of giant planets such as Jupiter, and in the early stages of fusion ignition. This regime presents a challenge for theory because there are no small parameters that facilitate approximations; for instance, in this state the thermal energies of electrons and ions are typically comparable to the Coulombic potential energy of interparticle interactions[1,2]. Therefore, to inspire and validate predictive models of this state of matter, new experimental results reporting unexpected phenomena are extremely valuable.

Our understanding of WDM has been greatly advanced by laboratory-based studies. Generating matter under extreme conditions on Earth is possible in large part thanks to laser facilities. High-intensity lasers are capable of reaching the peak powers necessary to produce the requisite temperatures and pressures in samples of interest[1,3]. Radiation from X-ray free electron lasers (XFELs) is particularly notable, as it has been used to both create and probe WDM. For example, XFELs have been employed to investigate nanoscopic diamonds[4,5] created through shock compression, produce high-density plasmas in silver[6], and study high-temperature high-pressure melting of aluminum[7], all of which involve transitions through WDM states[8]. Further, XFELs have been used to create and characterize highly ionized states of water under WDM conditions[9]. A detailed understanding of the behavior of water is particularly important for two reasons. First, due to its significance on Earth and anomalous properties, the structure of water has been extensively studied, both under ambient and extreme conditions[10-13]. Any new information about the structure of water can be placed in this context, driving towards a complete description of the water pressure-temperature phase diagram. Second, water is either a direct topic of study or an integral component of the sample in many experiments, including experiments performed at XFELs. Understanding the perturbative, damaging effects of XFEL radiation on water is necessary to properly design and interpret these experiments.

The aforementioned XFEL study on highly ionized water[9] was part of an experiment aimed at establishing whether or not specific radiation damage can be observed in protein crystals containing clusters of high-Z atoms[14]. To this end, data of water and protein microcrystals, respectively, were collected using short (25 fs) and unusually long (75 fs) XFEL pulses, with the latter chosen to maximize radiation damage effects. Damage induced by the XFEL pulse was observed, but the resulting dynamics were integrated over the duration of the pulse, preventing analysis of the temporal evolution of the damage process.

Therefore, we performed follow up X-ray pump X-ray probe studies at the Linac Coherent Light Source (LCLS) XFEL, allowing us to generate highly ionized states in both proteins and water and probe the resulting atomic structures with unprecedented resolution in time and space[15]. Here we describe our measurements on water, revealing that after exposure to a 7.1 keV X-ray pulse with a nominal fluence on the order of $5 \times 10^5$ J/cm$^2$, a previously undescribed structure of highly ionized water is formed. It is characterized by a peak in the wide-angle X-ray scattering (WAXS) profile at $q = 0.7$ Å$^{-1}$, corresponding to structural order at length scales of approximately 9 Å, i.e. significantly longer than the 2.8 Å and 4.5 Å oxygen-oxygen distances of the first and second solvation shells in liquid water under ambient conditions[16]. We performed molecular dynamics simulations, which predict that under these conditions the water sample is highly ionized, with two qualitatively distinct regimes. In the first regime, the high intensities found at center of the X-ray focus are sufficient to strip nearly all electrons from the sample. In the second lower intensity regime, outside the focal center, the irradiation produces singly and doubly ionized oxygen atoms, and the atomic structure rearranges significantly. The simulations, however, cannot account for the novel 9 Å structure observed, highlighting a gap in our theoretical toolbox or understanding.

## Results

**X-ray pump X-ray probe experiments reveal structural changes in highly ionized water**

To explore XFEL-induced radiation damage in protein crystals, we performed a two-color X-ray pump X-ray probe experiment at the Coherent X-ray Imaging (CXI)[17] endstation of the LCLS. During that experiment, we also investigated the effect of X-rays on liquid water, injected into the XFEL interaction region in the form of a ~5 μm continuously flowing column.

Per shot, the LCLS delivered ~1 mJ of X-ray energy which was split roughly equally between the pump and probe pulses, each approximately 15 fs in duration. To achieve high power densities, these pulses were focused with KB mirrors to a nominal 0.2 µm FWHM, corresponding to $3.5\times10^{12}$ photons/µm$^2$ or an average intensity of $2.7\times10^{19}$ W/cm$^2$ ($5\times10^5$ J/cm$^2$). Due to aberration, the focus is non-Gaussian in character, complicating the modelling of the intensity distribution. The focus contains a central spot of high intensity accompanied by "wings" that illuminate a larger total area, but with only a fraction of the pulse fluence[18] (*Materials and Methods*).

To separate the scattering patterns of the pump and probe, their photon energies were tuned to lie above and below the iron K-absorption edge (7.11 keV), respectively. A thin iron foil was placed in front of the detector, absorbing the pump but allowing the probe pulse to propagate[15,19] to a CSPAD area detector[20] (Supplementary Fig. 1 in reference[15]). The time delay between the pump and probe was tuned to between 20 and 110 fs. Due to significant jitter, the actual times as determined by the XTCAV diagnostic[21,22] covered a continuous range from 0 to >110 fs. In addition, we collected single-pulse (probe only) data as a reference. As the precision of the pulse arrival time measurements from the XTCAV are significantly less than the 15 fs pulse duration, we estimate the time resolution of our measurements to be 21 fs based on the convolution of two 15 fs pulses.

Visual inspection of the resulting 2D detector images showed clear and drastic changes of the scattering signal with increasing pump-probe time delays: the water ring signal decreased and appeared to migrate closer to the beam stop (Fig. 1A and 1B). This initial impression is supported by rigorous analysis of the data (Fig. 1C, *Materials and Methods*). With increasing time delay between the pump and probe pulse, we observed a significant decrease of the ambient water peak at $q = 2.1$ Å$^{-1}$ and the appearance of increased scattering between $q = 0.7$ and 1.5 Å$^{-1}$, that begins as a raised shoulder of the ambient water peak but resolves into a distinct new peak at $q = 0.7$ Å$^{-1}$ by 100 fs (the momentum transfer $q = (4\pi \sin \theta)/\lambda$, where $2\theta$ is the scattering angle and $\lambda$ is the wavelength of the incident X-ray beam). This new peak is broad, spanning from $q = 0.5$ to 0.9 Å$^{-1}$, implying a new short-range order in the sample. The characteristic length scale of $2\pi/q = 9$ Å is distinct from any known structure of water, leading us to assign this observation to a structural rearrangement of the ionized water sample. The relative magnitude of this peak increases as a function of pump pulse intensity, while the magnitude of the ambient water peak decreases (Fig. 1D and 1E). The newly formed structural arrangement persists until the end of our observation window at 110 fs pump-probe delay.

**Molecular dynamics simulations of the XFEL-water interaction establish two ionization regimes as a function of fluence**

To explain the origin and nature of the new structural arrangement we observed, we performed a set of calculations with the simulation tool XMDYN, a molecular-dynamics-based (MD) and Monte-Carlo-based code for modeling X-ray driven dynamics in complex systems[15,23-25]. XMDYN uses atomic cross sections calculated on-the-fly by the ab-inito code XATOM to capture X-ray atomic physics[23,26,27]. Our simulations modeled photoionization, Auger and fluorescence decays of core holes, and electron collisional ionization within X-ray irradiated water. Three-body recombination was not included, as it does not contribute significantly at <100 fs timescales.

While chemical bonds may be modeled by XMDYN using classical force fields[24,28] to mitigate computational cost, we proceeded under the assumption that chemical bonds can be neglected in our simulations. This assumption is valid if the ions move significantly less than a bond length before a majority of atoms are ionized at least once. After this point, Coulomb forces dominate the interparticle forces, and chemical bonds can be ignored. We established the time it takes to reach this fully ionized condition as a function of simulated X-ray fluence, discussed below. The XMDYN model therefore captures Coulomb interactions between charged particles but does not treat interactions between neutral atoms or between neutral atoms and ions.

As it was prohibitively expensive to simulate all atoms and electrons in the irradiated 0.2 μm FWHM section of a 5 μm diameter water jet, we restricted the simulations to cubic volumes with an edge length of 47 Å and employed periodic boundary conditions[29,30]. This box size was selected to ensure a statistically relevant number of photoionization events in each simulation and mitigate periodic boundary artifacts, following tests with cubes with edge lengths of 30, 47, and 60 Å (*Methods and Materials*, Fig. S1, Supplementary Table 1). For the simulations time zero is defined as the maximum of the pump pulse.

As in our previous X-ray pump X-ray probe investigation on protein nanocrystals[15], we performed simulations at multiple fluences to both hedge against absolute uncertainty in the focal spot intensity and provide information on the ionization dynamics within the water jet at different distances from the X-ray focus center. The X-ray focus size was characterized by imprints, which have limited precision (*Materials and Methods*). Further, the X-ray focus does not have an ideal Gaussian spatial profile. A significant part of the pulse energy is deposited in the "wings" of the focused beam. Since the water jet is much larger than the focused X-ray beam, these out-of-focus regions contribute appreciably to the scattering signal, but have been pumped with a much lower fluence than the center of the focus. Therefore, we simulated irradiation by X-ray pulses of with fluence values ranging from 1% to 100% of the nominal experimental value (Fig. S2).

We employed this fluence titration to test the validity of our model, which assumes high levels of ionization ensure Coulombic interactions dominate, such that covalent bonds and intermolecular forces can be considered negligible. At 1% of the nominal experimental fluence, only 25% of the simulated oxygen atoms are ionized at least once by 110 fs. In this time, the $O^{1+}$ ions move 3.1 Å on average, more than 3 times the O-H bond length in ambient water. Therefore, we expect covalent bonds would have a strong influence on the results, and their neglect is not justified. At 2.5% fluence, the average displacement of $O^{1+}$ ions reaches about 1 Å (the typical O-H bond length) at the time of about 50 fs after the pump pulse. At this time, 25% of atoms are still neutral, and bonds are still expected to influence the dynamics. By 10% of the nominal fluence, however, covalent bonds are expected to play a negligible role, as there are no neutral atoms left at the end of the 15 fs FWHM pump pulse, in which time and the ions are displaced by <1 Å. Moreover, in a simulated cube with 47 Å edge length, 10% fluence irradiation yields 48 primary photoionization events, a large enough number that we expect this simulation to be representative of the bulk. Therefore, we conclude that our model assumptions are valid for fluences of 10% and greater.

Inspecting the results of this fluence titration between 10% and 100% of the nominal fluence revealed two qualitatively different regimes of ionization dynamics, with the first spanning from 10% up to ~20% and the second appearing at >20% of the nominal experimental fluence. Both regimes are expected to contribute to the measured experimental signal, which is produced from a superposition of scattering from the high-fluence focal center and weaker beam wings (Fig. 2A). Given the difficulties in characterizing the experimental focus and uncertainty in the experimental fluence (*Materials and Methods*), we do not make quantitative predictions of the experimental signals using the simulations but employ these two qualitative regimes predicted by XMDYN to gain insight into the states of matter generated and corresponding scattering signals we would expect. We found that the 100% and 10% fluence simulations were good representative examples of the two qualitative fluence regimes, and these simulations are presented in detail in the main text; additional simulations are presented in the Supplementary Figures.

**Distinct ionization mechanisms at high and low pump pulse intensity result in specific scattering curves**

The two fluence regimes are characterized by distinct dynamics and scattering curves. At high fluence (100%), a significant fraction of ions are generated by primary photoionization (Fig. 2B). Approximately 15% of the atoms in the sample are ionized at least once through a direct interaction with the pump photons. Subsequent Auger decays and secondary electron impact ionization create highly charged states, ultimately resulting in a sample consisting primarily of $O^{5+}$ and $O^{6+}$ ions (Fig. 2F). While primary ionization is prevalent, secondary ionization still plays a dominant role, as on average a single photoionization event results in 25 secondary ionizations. At high fluence, the amplitude of the scattering curve $I(q)$ decreases dramatically at all scattering angles but remains essentially unchanged in shape (Fig. 2H). The water sample becomes strongly ionized, and as the majority of bound electrons are stripped from their associated atomic nuclei the oxygen form factors are attenuated, as observed by Inoue et al.[31] and reported previously[9].

At 10% fluence, only 1.5% of atoms in the sample undergo primary photoionization by the pump. Under this weaker irradiation, secondary ionization cascades play a more significant role (Fig. 2C). On average, a single photoionization is followed by 144 secondary ionization events. The highest charge states reached are $O^{1+}$ to $O^{3+}$, with electrons ejected primarily from the valence shell by electron impact events (Fig. 2G). As in the high-fluence case, the scattered intensity $I(q)$ shows a marked decrease in the ambient water peak as a function of time. However, in contrast to the high-fluence regime, this is accompanied by an increase in scattering intensity at lower scattering angles ($q < 1.7$ Å$^{-1}$, Fig. 2I). By ~70 fs, $I(q)$ is nearly flat as a function of $q$. This "white" curve suggests structure on many length scales, with no characteristic inter-ion or inter-electron separation. The predicted decrease in the ambient water peak with simultaneous increase in scattered intensity at low-$q$ values is consistent with the experimental data. However, although the simulations show an increase of scattering signal at low scattering angles, they do not show the emerging peak at $q = 0.7$ Å$^{-1}$ observed in the experiment.

After ~20 fs, simulations predict the irradiated water sample in both the 100% and 10% fluence simulations enters the WDM regime (Fig. 3). The kinetic electron temperatures stabilize at ~19 eV under 10% fluence irradiation and ~120 eV at 100% fluence, with free electron number densities on the order of $10^{23}$ cm$^{-3}$. In contrast, the kinetic temperature of the pumped oxygen ions is 0.4 eV at 10% fluence case and 2 eV at 100% fluence at 20 fs i.e., far below the respective electronic temperatures. The ion temperatures in both the high and low fluence simulations continue to rise as a function of pump-probe time delay. By the end of the simulations at 110 fs, the X-rays have generated unthermalized WDM, in which the electron-ion system is still far from equilibrium.

**Ion dynamics account for changes in the scattering curve at 10% fluence**

The pump-induced ionization of the water sample has two consequences that result in changes to the scattered X-ray intensity: the atomic form factors[32,33] attenuate and, over time, the spatial arrangement of the oxygen and hydrogen ions (resulting from ionized water molecules) changes. Importantly, our simulations enabled us to predict the impact on the X-ray scattering caused by these effects and separate them from one another.

This allowed us to distinguish between two possible mechanisms for the changes in $I(q)$ observed at 10% fluence in simulation. In one model, specific patterns of ionization could occur, producing structure in the sample that is reflected in the scattering curve. Alternatively, atomic rearrangements induced by Coulomb forces following ionization could produce the new structure. Combinations of these effects are also possible. To distinguish between these possibilities, we computed X-ray scattering curves with: (i) atomic displacements accounted for but using neutral O$^{+0}$ form factors for all atoms; (ii) only form factor changes, with ions fixed in their initial positions; (iii) both atomic displacements and form factor changes. The results are shown in Fig. 4 for water irradiated with 10 % and 100 % nominal fluence, respectively, at a 110 fs pump-probe delay. The scattering from unirradiated water is shown for comparison.

This analysis shows that in the case of 10% fluence, ion motions are the primary contributor to the change in the shape of the scattering curve. Ionization results in a "randomization" of the original hydrogen bonded structure, with charges created non-uniformly throughout the irradiated volume. Following ionization, Coulomb forces induce atomic motion that further disrupts the ambient water structure. The more highly charged $O^{2+}$ ions undergo larger displacements than $O^{1+}$ ions, moving on average 7.0 Å and 5.6 Å, respectively, within the first 110 fs of the simulation. In contrast, at 100% fluence, the pump radiation causes a very strong reduction of atomic form factors, significantly decreasing the scattering from regions irradiated with the maximal pulse fluence (Fig. 4). This is the primary reason for the predicted changes in $I(q)$; atomic displacements have little effect in this case.

## Discussion

In our time-resolved X-ray pump X-ray probe solution scattering experiment on liquid water, we observe fluence- and time-dependent changes of the scattering curves compared to those of undamaged water. Specifically, the magnitude of the ambient water peak decreases, and a new distinct low-$q$ feature appears at longer pump-probe time delays, corresponding to a previously undescribed feature of highly ionized water.

The XFEL pulse contains an intense center of the focused beam surrounded by much weaker and larger halo or "wings". Both contribute to the observed scattering. Simulations of the experiment suggest this signal, integrated over many pump fluences, can be characterized by two qualitative regimes. At the highest power densities, the majority of bound electrons are stripped from the water molecules, and the resulting scattered intensity of the probe is greatly reduced, bleaching the signal. At low (fluences ≤ 20% of the nominal maximum, Fig. S2) the sample is less strongly ionized and the ions in these regions re-order over ~75 fs into a new configuration, disrupting the spatial correlations present in neutral water. While our simulations do predict an increase in low-$q$ scattering intensity caused by structural rearrangements, they do not reproduce a distinct new peak at $q = 0.7$ Å$^{-1}$.

Both our experimental and simulation findings differ from the ones described in a previous study of highly ionized water conducted at the LCLS[9]. In that study, solution scattering data from XFEL pulses of 25 and 75 fs duration, integrated over the entire pulse, were compared to each other and to data of presumably undamaged water collected at a synchrotron beamline[9]. Upon XFEL irradiation, the ambient water peak shifted to slightly lower scattering angles. For long pulses, the peak broadened. As these changes are qualitatively different from what we observe, we sought points of comparison that might explain why the structural change we report was not observed in that experiment. First, the ionization dynamics that occur during continuous irradiation and during pulsed irradiation followed by a reaction period differ, and the observed signal is distinct due to the temporal integration over the single long pulse (Fig. S3). Second, the peak of the low-$q$ feature we report was outside the range of scattering angles reported in ref. [9], which reports scattering from $q = 0.96$ Å$^{-1}$ to 2.6 Å$^{-1}$ (using our convention). Unnormalized versions of the $I(q)$ curves from ref. [9] do show a small rise in scattered intensity at the smallest $q$-values measured (Kenneth Beyerlein, personal communication), consistent with our observations reported here. Third, the nominal fluence employed in the previous experiment[9] was estimated to be $12.3 \times 10^{12}$ ph/µm$^2$ (6.86 keV photon energy, $>10^6$ J/cm$^2$), a few times higher than we estimated for the current experiment (*Materials and Methods*), though the average pulse powers are similar ($3 \times 10^{19}$ W/cm$^2$ here vs. $5.4 \times 10^{19}$ W/cm$^2$ and $1.8 \times 10^{19}$ W/cm$^2$ for the 25 fs and 75 fs pulses in ref [9], respectively). The actual power densities employed in the two experiments may have differed significantly from these nominal values due to inherent limitations on the ability to reproducibly achieve optimal focus at CXI (see *Materials and Methods*). Therefore, while both experiments report X-ray solution scattering from highly ionized water, key differences in the experimental parameters result in qualitatively different findings.

Particularly striking is our observation of a distinct low-$q$ peak that emerges at pump-probe time delays ≥75 fs. This feature is reminiscent of a low-$q$ peak detected in solutions of Mg$^{2+}$, Al$^{3+}$, Ni$^{2+}$ and Fe$^{3+}$ ions in water[34-36]. At concentrations on the order of one molar, the tight hydration shells around these ions minimize the inter-ion distance and result in a large density contrast in elastic scattering data[36]. Since our simulations predict ionization of essentially all oxygen atoms upon exposure of the water sample to 10% of the nominal fluence, it is conceivable that our low-$q$ peak originates from water exposed to even lower X-ray fluences, for instance farther outside the focus. The lower fluence in these regions could produce oxygen ions at concentrations on the order of one molar solvated by undamaged water. A computational analysis in this fluence regime is complicated by the fact that no general modeling description exists for bulk water at arbitrary ionization degree.

The significant perturbation of liquid water we observed has strong implications for other experiments that rely on the assumption that an XFEL pulse is effectively non-perturbative. Especially notable examples are experiments involving the interaction of two subsequent XFEL pulses with a single sample. This includes X-ray photon correlation spectroscopy (XPCS) experiments, an X-ray probe X-ray probe technique, for which water has been a sample of interest[37-42] (see Supplementary Table 1 for details). An underlying assumption of these experiments is that any probed structural changes are due to dynamics of interest, not perturbations induced by the first X-ray pulse. Another example are experiments where the first X-ray pulse is used to photoreduce the sample, an approach that has been used at synchrotron sources[43] and is being discussed for XFEL applications. Our results show that the structure of even neat water is strongly perturbed when the absorbed dose results in formation a significant number of ions and enough time passes to allow reordering of their solvation shells. Thus, two-XFEL-pulse experiments that aim for unperturbed measurement conditions should check parameters affecting the dose (photon energy, here 7.11 keV; fluence, here nominally $\sim 10^{11}$ photons/$\mu m^2$) and thus number of generated ions as well as the time delay between pump and probe pulse.

While at synchrotrons the dangers of high fluence (and thus dose) on sample integrity are well known, our study suggests that at XFELs due to the high dose rate even "low" fluences may disrupt the structure of water. This damage is only observable after sufficiently long time delays (here $\geq$ 75 fs) when the generated ion concentrations are in the molar range. Our study does not allow us to give an experimentally derived estimate for a "safe" fluence/dose below which this structural rearrangement is negligible; follow up studies are needed to do so. However, the previously discussed studies of solutions of divalent and trivalent cations show that molar concentrations of ions are sufficient to significantly rearrange the structure of water[34-36] and produce a scattering peak at low $q$ similar to the one we observe. Our simulations predict that the concentration of $O^{2+}$ would reach molar concentrations within 110 fs after exposure to a pulse with only 1% of our nominal fluence (3.5×10$^{10}$ photons/μm$^2$ or an average intensity of 2.7×10$^{17}$ W/cm$^2$). Since this X-ray intensity is quite low for a standard XFEL experiment, it is highly advisable for two-pulse XFEL studies to perform a fluence titration, monitoring WAXS signals with delays ~100 fs and longer. These measurements can establish the unperturbed regime and ensure the material under study is the expected one. Our results demonstrate that such unexpected states of matter can be readily generated by the high peak powers of XFEL radiation, presenting both challenges and opportunities for experiments with XFEL light.

## Conclusion

Using an X-ray pump X-ray probe experimental setup, we report an unexpected structural change in ionized water under WDM conditions, furthering our understanding of the properties this important liquid. Our key experimental observations are the attenuation of the ambient water peak and formation of a new low-$q$ peak at ~75 fs after ionizing pump irradiation, corresponding to a previously unobserved structural rearrangement of ionized water. Our study shows that even low fluence XFEL irradiation (~10$^{11}$ photons/μm$^2$ at 7.11 keV) can cause significant changes in the water structure for time delays exceeding 75 fs and predict that this also occurs at much lower fluences. Because water plays a role in many XFEL-based experiments where X-ray induced perturbations may interfere with the interpretation of the primary scientific aim of the study, this is essential information for the future design and interpretation of such experiments.

Simulations predict that due to the non-uniformity of the beam focus, two different regimes of ionized water are generated. In the focal center, the sample can be highly ionized and as a result becomes effectively transparent to the pump. In contrast, in areas illuminated by the less intense wings of the XFEL beam, the structure of water changes resulting in a change to the observed scattering curve. This model is able to describe the attenuation of the ambient water peak and a rise in low-$q$ scattering, but fails to account for the new order at $q = 0.7$ Å$^{-1}$. The observation of peaks at similar scattering angles in solutions of divalent and trivalent cations suggests that this feature may be due to the generation and subsequent solvation of oxygen ions. This remains to be proven, however, and a microscopic description of the structure of the novel structural rearrangement we report remains an open challenge for theory that, if solved, can advance our understanding of water under extreme conditions.

**Material and methods**

**Experiment.** The experiment was performed in the nanofocus chamber of the Coherent X-ray Imaging (CXI) instrument[17] at the Linac Coherent Light Source (LCLS) in February 2015 (proposal LG07/LE70). In between injections of protein nanocrystals (results were reported previously[15]) we introduced water into the XFEL beam in a ~5 µm cylindrical liquid column using a gas dynamic virtual (GDVN) nozzle injector[44]. The position of the sample jet was continuously adjusted to maximize the hit rate. To follow the time-dependent X-ray-induced dynamics, an X-ray pump X-ray probe scheme was used[19] as shown in Supplementary Fig. 1a in reference [15]. Two ~15 fs X-ray pulses were produced using the double-pulse operating mode at the LCLS[45], with the pulses delayed by 0 to 120 fs in time with respect to one another. The pump pulse was tuned ~40 eV above the iron K-edge at 7.11 keV, while the probe pulse was ~40 eV below the edge[15,19]. Instabilities in the FEL generated jitter in pump-probe delay, which was monitored by the XTCAV[21,22], and in the pulse energies, which were monitored by the a diffractive spectrometer in the LCLS Front-End Enclosure (FEE)[46]. Both the XTCAV and FEE spectrometer provide information on a single-shot basis. An iron foil was used to calibrate the FEE spectrometer; an energy sweep enabled us to precisely locate the Fe K-edge for later analysis. A 25 µm Fe foil placed downstream of the sample effectively blocked scattering from the pump pulse but allowed the probe to propagate to a Cornell-SLAC pixel array detector (CSPAD)[20] ~70 mm downstream of the interaction region. As a control, we also took data without the pump pulse, providing an unperturbed reference signal.

The data was collected in two shifts of 24 and 36 hours, respectively. At the beginning of each shift, the X-ray focus was optimized using imprints, a method by which the beam profile is deduced from the size of a vaporized area on a thin gold film hit by the beam at various intensity levels[47]. The diameter of the X-ray focus was determined to be ~0.2 µm FWHM. With a beamline transmission of ~ 45% the power density at the interaction region was nominally $2.7 \times 10^{19}$ W/cm$^2$ (corresponding to $3.5 \times 10^{12}$ photons/µm$^2$ per single pulse). This estimate provides an upper bound; the actual power density was likely lower.

Ronchi shearing interferometry performed in a separate experiment after ours showed that the CXI focus consists of a central focal region with strong "wings" that contain 10% to 50% of the intensity of the central spot at optimal focus[18]. However, as discussed in the supplementary information of ref [15], significant uncertainties in the actual focus used in our experiment exists. The imprint method employed during the experiment[47], while state of the art at the time, provides inherently limited precision in the ability to align the beamline optics and achieve optimal focus. Any imperfect alignment of the KB mirrors, aberrations of the mirror surfaces, misalignment of the water jet with the focal plane, or jitter in the spatial trajectory of the XFEL pulses will all produce a less tight focus than is theoretically achievable. Finally, thermal or mechanical drift during data collection and uncertainties in the transmission of the beamline contribute to our overall uncertainty in the absolute fluence values in the focus. Therefore, we report nominal fluence values representing our best estimates, but acknowledge that due to experimental realities these values provide an upper bound on the actual fluences that were effective during the experiment. The same considerations concerning the flux density apply to the previous experiment analyzed by Beyerlein et al[9].

**Data Analysis.** X-ray images captured on the CSPAD were processed using *psana*[48] and custom python code (https://github.com/tjlane/cxig0715). After dark subtraction, gain correction, and pixel masking, images were averaged over the azimuthal angle into 500 radial bins, with $\Delta q \approx 0.00745$ Å$^{-1}$ per bin, to produce $I(q)$ curves for each X-ray pump/probe event. FEL pulses are generated from a stochastic process, causing jitter in the per-pulse photon energy, arrival time, and spatial trajectory of each shot. Therefore, images were sorted to ensure they met specific experimental criteria before inclusion in downstream analysis. Specifically, images were analyzed only if the XFEL intersected the liquid jet, producing a liquid $I(q)$ trace with no identifiable protein crystal Bragg peaks as judged by a custom peak-finding algorithm (hit finding). Further, images were rejected if the pump and probe pulses were not fully above and below the iron K-edge, as monitored by the FEE spectrometer[46]. For retained shots, relative X-ray pump/probe arrival times were determined by the XTCAV instrument[21,22] and sorted into 10 fs bins from 0-10 fs to 100-110 fs. Data in these bins were normalized per-pulse by the total measured intensity in the probe as measured by the FEE spectrometer and averaged to produce the presented $I(q)$ traces.

The final traces showed up to twofold variation in total intensity from timepoint-to-timepoint. This was attributed to imperfectly corrected variation in integrated pulse energy, as any change in pump-probe delay required reconfiguration of the LCLS operating mode. In the absence of accurate diagnostics to correct for this variation, we normalized our experimental curves by the total scattered intensity,

$$I_{norm}[q] = I[q]/\sum_q I[q]$$

where the sum spans discrete bins covering the $q$-range from 0.37 to 2.52 Å$^{-1}$, the extent of the presented data.

**Simulation**

**Initialization of the water simulation.** XMDYN simulations were initialized from short classical simulations performed with NAMD2[49]. Briefly, a cubic box with periodic boundary conditions was filled with water modeled by the TIP4P force field using VMD[50]. NVT equilibration was performed at 300 K using a Langevin thermostat. Temperatures were monitored, and once converged, the final atomic positions were used as the initial condition for XMDYN simulations.

**XMDYN simulations of XFEL-water interaction.** XMDYN[23,24] simulations were performed to model the dynamics of the system under intense X-ray irradiation, including the processes of photoionization, Auger and fluorescence decays of core holes, and electron collisional ionization. The system was treated using periodic boundary conditions. An X-ray pump pulse of 7.11 keV photons with a Gaussian temporal profile (15 fs FWHM) was introduced in a spatially uniform fashion across the sample. Fluence levels ranged from 1% to 100% of the nominal experimental value of 3.5×10$^{12}$ photons/µm$^2$. Simulations were begun 30 fs before the peak of the X-ray pulse and propagated to 110 fs after this peak. We conducted simulations with and without a probe pulse and for various box sizes, as detailed below.

**Optimization of the periodic box size.** Periodic boundary conditions can introduce artifacts in the diffraction signal at low $q$ values, close to the reciprocal of the box size. Therefore, we tested different box sizes to balance computational expense with the accessible $q$-range. Figure S1 show the behavior of azimuthally averaged diffraction signal at 10% and 100% nominal fluence for cubic boxes with edge lengths of 30 Å, 47 Å, and 60 Å, corresponding to $q$ = 0.21 Å$^{-1}$, 0.13 Å$^{-1}$, and 0.10 Å$^{-1}$. The diffraction signal was calculated at the time point of the probe pulse maximum. As a low $q$ (long range) artifact was observed for the 30 Å box size, but not the others, we selected a 47 Å box for future simulations.

**Computation of diffraction intensity $I(q)$.** I(q) was computed by summing the scattering from all particles in the simulation,

$$I(\boldsymbol{q},t) \propto \left| \sum_k f_k(q,t) e^{-i\boldsymbol{q}\cdot\boldsymbol{R}_{k(t)}} \right|^2$$

where atomic scattering form factor, $f_k(q,t)$, for each atomic configuration was computed by the XATOM code[23,26,27]. Here, $t$ represents the timestep of the simulation and $\mathbf{R}_k(t)$ the position of particle $k$. Reciprocal lattice vectors $\mathbf{q}$ were sampled randomly in the reciprocal space and then averaged in annular bins.

**Simulated effect of the probe on the structure of pumped water.** We performed simulations to understand the impact of the probe pulse on the evolution of the sample. We tested fluences of 10%, 50%, and 100% of the nominal experimental fluence and observed only small effects of the probe pulse on the molecular dynamics and resulting $I(q)$ curves. The probe interacts with a system that has already been strongly ionized by the pump, and therefore has a severely attenuated effect on the sample compared to the pump pulse, which interacts with neutral water. Consequently, we assumed that the perturbative effect of the probe was negligible and did not simulate all pump-probe delay scenarios explicitly but analyzed the temporal evolution of pump-only simulations. This reduced the number of computationally expensive simulations substantially, as we did not simulate each pump-probe delay separately.

**Data Availability**

Experimentally and computationally derived scattering intensities $I(q,t)$ have been deposited in the github repository https://github.com/tjlane/cxig0715. The repository will be made publicly available upon acceptance of the manuscript.

**Code Availability**

Analysis scripts can be retrieved from https://github.com/tjlane/cxig0715. The repository will be made publicly available upon acceptance of the manuscript.

**Acknowledgements**

I.S. thanks Alexander Kozlov and Harry Quiney for contributions at early stages of the project. We thank the Heidelberg FEL group and the LCLS CXI team for collecting the water data in February 2015. We thank Anders Nilsson, Foivos Perakis and Kenneth Beyerlein for stimulating discussions. A.P.M. and B.Z. gratefully acknowledge the funding received from R & D grant of the European XFEL, with the contribution of IFJ PAN in Krakow. M.S. is on leave from the Institute of Physics, Czech Academy of Sciences, Na Slovance 2, 182 21 Prague 8, Czech Republic. TJL was supported by a Helmholtz Young Investigator Group (YIG) award. Z.J. acknowledges support from DESY (Germany), a member of the Helmholtz Association HGF. We acknowledge financial support obtained from the Cluster of Excellence 'Advanced Imaging of Matter' of the Deutsche Forschungsgemeinschaft (DFG) - EXC 2056 - project ID 390715994. Use of the Linac Coherent Light Source (LCLS), SLAC National Accelerator Laboratory, is supported by the U.S. Department of Energy, Office of Science, Office of Basic Energy Sciences under Contract No. DE-AC02-76SF00515. Parts of the sample injector used at LCLS for this research was funded by the National Institutes of Health, P41GM103393, formerly P41RR001209.

**Author Contributions**

S.B. conceived the experiment which was designed and coordinated by S.B. and I.S.; A.G. performed online analysis and the initial offline data analysis; T.J.L. performed the final data analysis; Z.J., A.P.M., M.S., B.Z. developed the XMDYN-based modelling strategy; M.S. performed the XMDYN calculations; Z.J. and B.Z. supervised the calculations. All authors discussed the results and contributed to the manuscript. The initial versions of the manuscript were written by T.J.L., I.S., and B.Z.

**Competing interests**

The authors declare no competing interests.

**Supplementary Information is available for this manuscript.**

Correspondence and requests for materials should be addressed to TJL, IS, BZ

**Figures and Figure Legends**

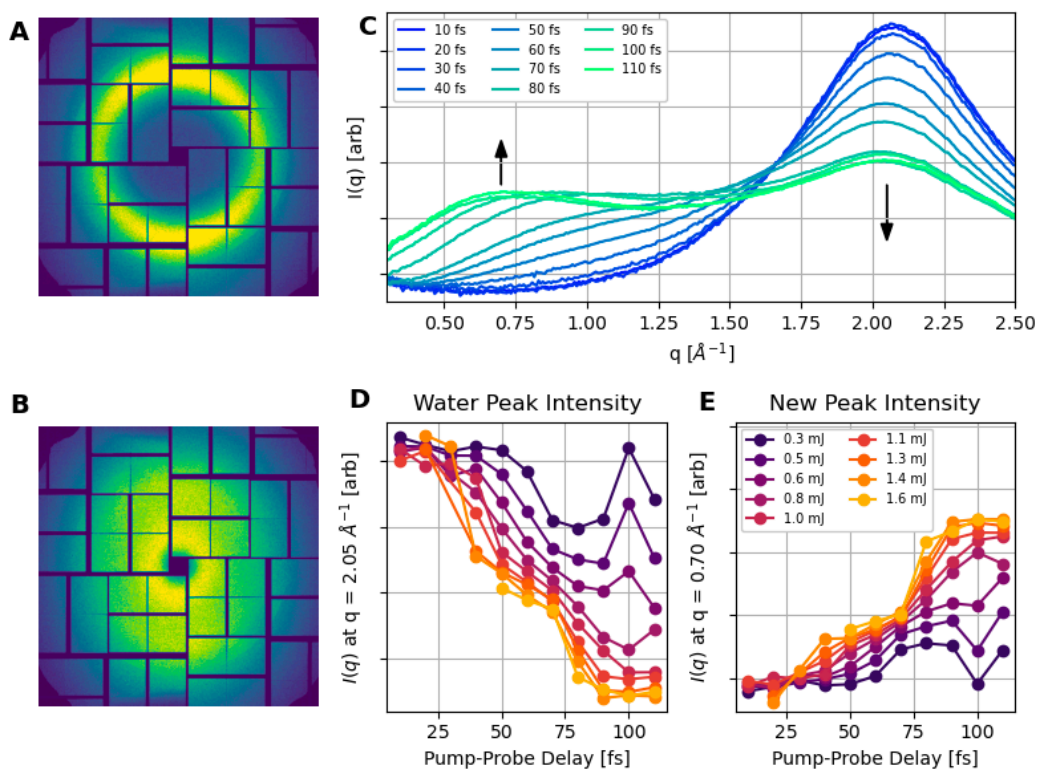

Figure 1. X-ray pump X-ray probe experiments reveal a new structural change in ionized water. (**A**) X-ray probe only data show a single diffraction ring corresponding to the unperturbed structure of liquid water, while (**B**) X-ray pump X-ray probe data (100 fs delay) show a distinct new diffraction ring at low scattering angles corresponding to a , previously undescribed structural change. Both images show an average of 10 pulses from the LCLS captured on the CSPAD detector. (**C**) This new structure forms within ~75 fs and is characterized by a peak in $I(q)$ at $q = 0.7$ Å$^{-1}$ (upward arrow). We observe concurrent attenuation of the ambient water peak at q = 2.05 Å$^{-1}$ (downward arrow). Curves were normalized by total scattering intensity for the $q$-range shown (0.37 to 2.52 Å$^{-1}$, see *Materials and Methods*). Sorting of the pump pulses by fluence reveals this response is fluence-dependent (*Materials and Methods*). (**D**) As a function of pump fluence, the ambient water peak height decreases, while (**E**) the intensity of the new peak increases concomitantly. For panels (D) and (E), $I(q)$ curves were normalized as previously described; peak heights are relative.

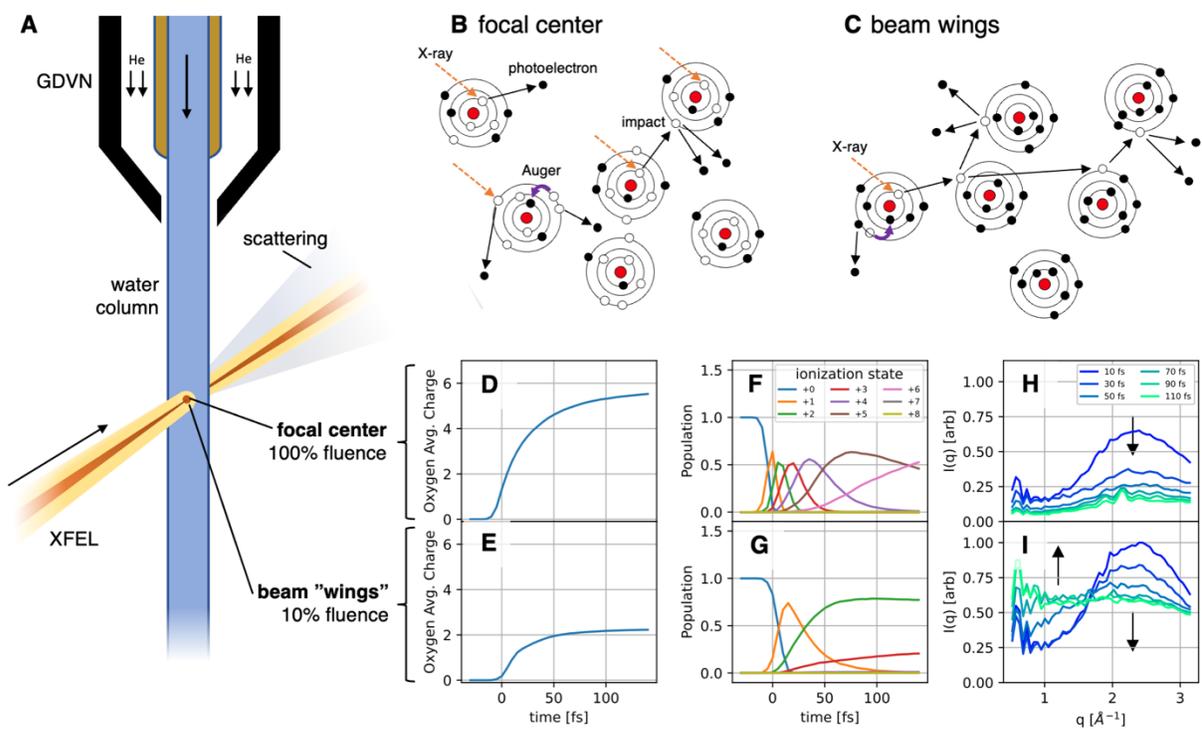

**Figure 2. Varying the pump fluence leads to qualitatively different ionization processes in XMDYN simulations.** (**A**) The final signal is a superposition of scattering from an intense focus and more weakly irradiated out-of-focus "wings" of the XFEL beam. These regions were characterized by simulations conducted at pump fluences of 100% and 10% of the nominal experimental value, respectively. (**B**) In simulations of the high pump fluence regime, 15% of atoms undergo primary photoionization, typically followed by Auger decay. Secondary ionization is significant, with each primary event generating 25 secondary ions through valence shell impact events. (**D**, **F**) These ionization events rapidly result in an average oxygen ion charge state of >5+ and production of oxygen ions with up to 6+ charge. (**C**) At lower pump fluence, primary ionization is greatly reduced, affecting only 1.5% of all atoms. Because the surrounding atoms still have most of their bound electrons, the cross section for secondary impact ionization is significantly higher than in regime (**B**), with 144 secondaries generated per primary photoionization. (**E**, **G**) This produces significant quantities of $O^{2+}$ ions and an average oxygen charge state of approximately 2+. The final observed scatter is expected to contain contributions from both of these regimes. (**H**) Upon high fluence irradiation the water scattering (B) is effectively bleached due to the ionization of scattering electrons, and shows no new structure. (**I**) under less intense irradiation in regime (**C**), the low-*q* scatter is predicted to rise while the ambient water peak is attenuated, similar to what is observed experimentally. No distinct new peak at low-*q* matching the experimental observation (Fig. 1C) is observed, however. Panels (B) and (C) are illustrations only and are not quantitative. Time zero corresponds to the intensity maximum of the pump pulse.

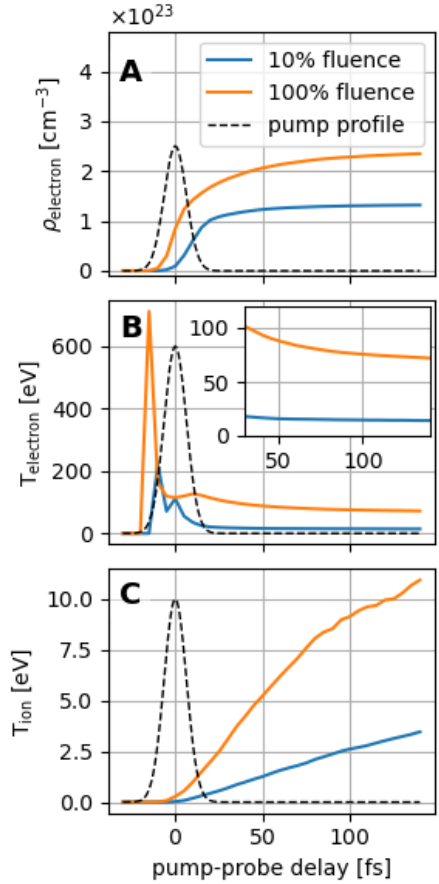

**Figure 3. Simulations predict a nonequilibrium state in the warm dense matter regime.** (**A**) After exposure to the X-ray pump significant number of free electrons, with a density on the order of $10^{23}$ cm$^{-3}$, are generated ($\rho_{electron}$). Shown are simulations for pump pulses with both 10% (blue) and 100% (orange) of the nominal experimental fluence, along with the temporal profile of the pump (black dashed line, height arbitrary). (**B**) The kinetic energy of the electron subsystem stabilizes within the duration of the pump. Shown is the kinetic energy as a temperature, $T_{electron}$, where $\frac{2}{3}kT = \langle \frac{1}{2}mv^2 \rangle$; the electrons stabilize at energies of approximately 20 eV and 75 eV for the 10% and 100% fluence pumps respectively. The large spikes at short timescales originate from the fact that only a few, high-energy free electrons contribute to the average at short timescales. Insert is the same data, rescaled to show the values at long time delays more clearly. (**C**) The ion subsystem heats continuously after irradiation by the pump. At 20 fs delay, the kinetic temperatures of the ions are approximately 0.4 eV and 2 eV for the 10% and 100% fluence cases, significantly lower than the electron subsystem. These temperatures continue to increase for the duration of the simulation, up to 110 fs after the pump.

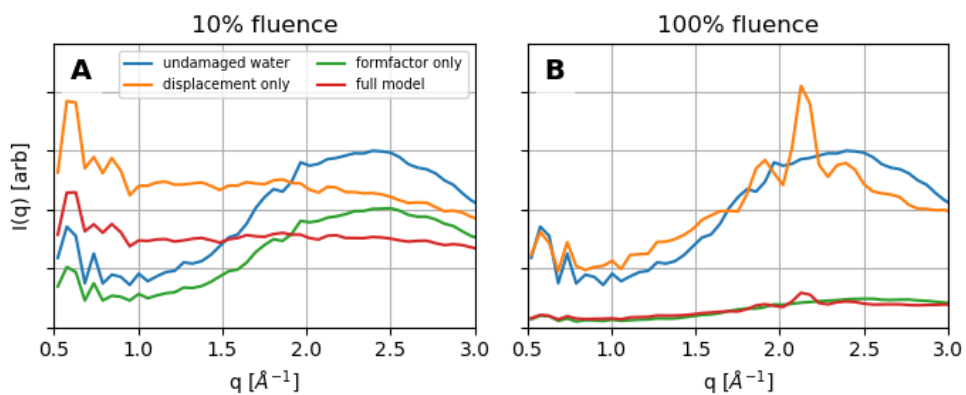

**Figure 4. Ion motion accounts for the change in shape of *I(q)* at 10% fluence, while the 100% fluence case is dominated by a reduction in form factor amplitude due to ionization.** Simulations of *I(q)* accounting for the effects of changes in atomic form factor (green), ion displacement (orange), or both (red) reveal that at (**A**) 10% fluence, ion displacement is the primary effect driving changes in the shape of *I(q)*, while (**B**) at 100% fluence, ion motion plays a minor role. In contrast, changes in the form factors due to severe ionization are the primary effect resulting in the observed attenuation of *I(q)*. The predicted scattering for simulated undamaged water (blue) is shown as a reference.

# Supplementary Information

*Stransky et al., Distinct transient structural rearrangement of ionized water revealed by XFEL X-ray pump X-ray probe experiment*

Supplementary Figures (1-3)

Supplementary Tables (1-2)

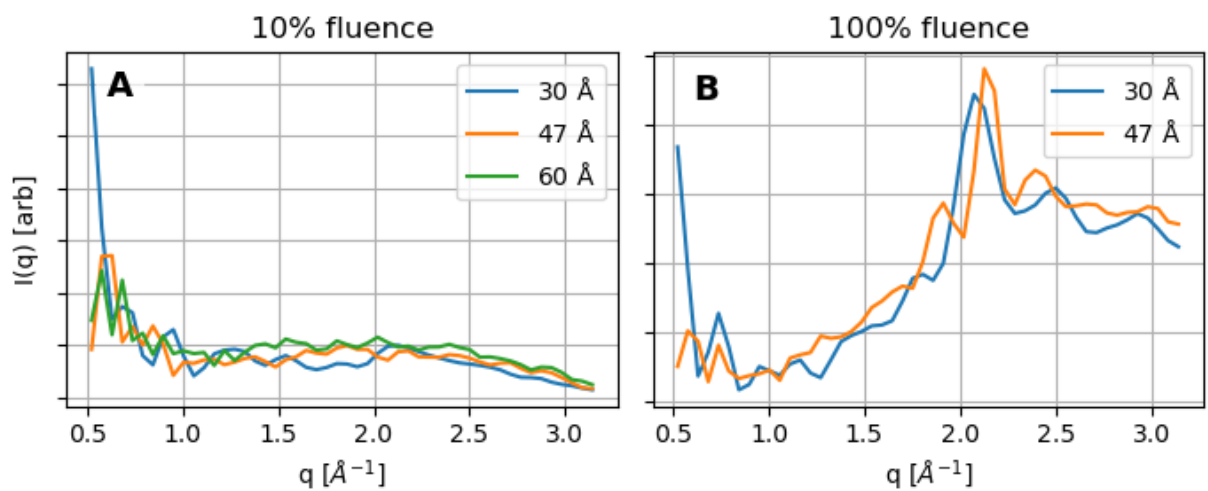

**Supplementary Figure S1**. **Finite size effects introduced by specific size of the simulation box.** Convergence of the simulated diffraction signal at (A) 10% and (B) 100% nominal fluence calculated at the time point of the probe pulse maximum for the indicated edge length of the water box.

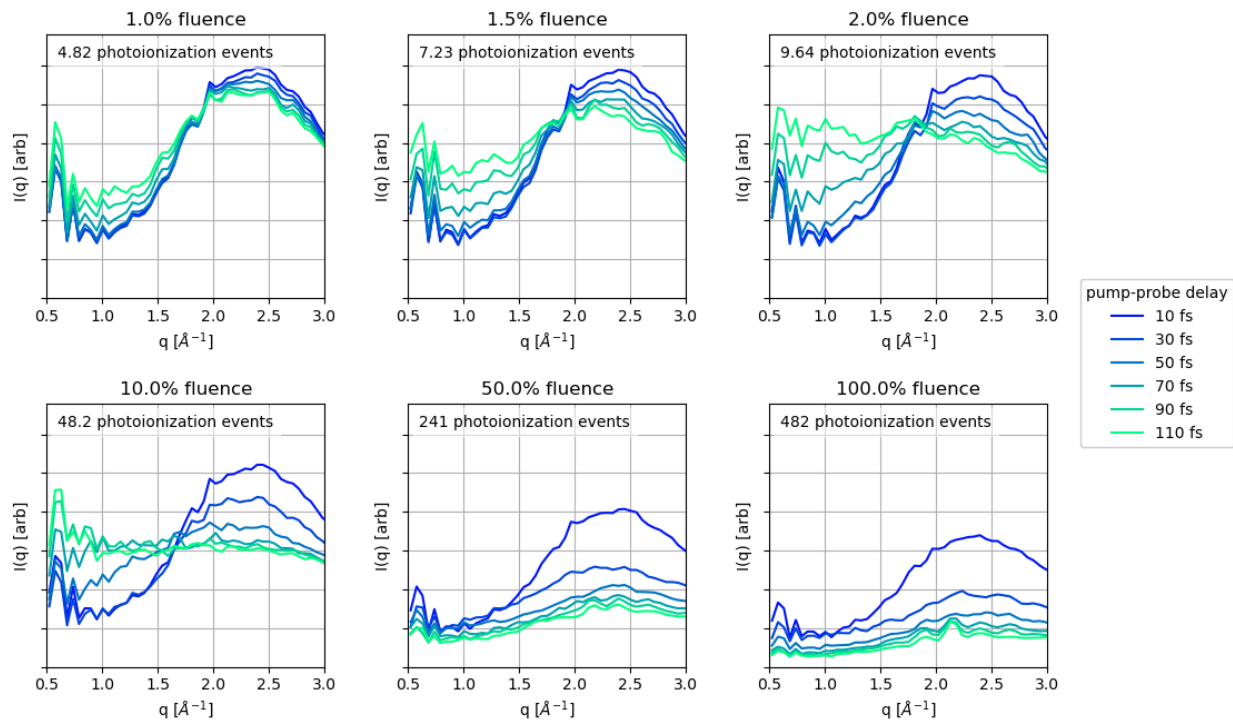

**Supplementary Figure S2. Fluence titration establishes two qualitative regimes of ionization.** Shown are computed $I(q)$ scattering curves as a function of time delay for different nominal pump fluences. Perturbation effects due to the X-ray interaction, such as the low-$q$ features in the scattering patterns, are seen at pump intensities as low as 1.5% of the experimental fluence. Between 2% and 10%, a flattening of the I($q$) curve is apparent, with a decrease in intensity of the ambient water signal and an increase in low-$q$ scattering. By 50% fluence, the signal is strongly attenuated at all scattering angles. This established two qualitatively different regimes, approximately spanning below and above 20% of the nominal fluence. Further, this power titration allowed us to establish the range of validity of our approach. The average number of primary photoionization events are indicated for each fluence. Due to the low number of photoionization events at low fluence (less than 10%, see Supplementary Table 1) a large fraction of chemical bonds are maintained. Because these bonds were not included in our model (see main text), this limits the predictive power of these simulations at fluences <10%. Further, a low number of primary photoionization events at these fluences mean each small simulation box is a poor approximation for a bulk volume of material, limiting the simulation's statistical power. Therefore we restricted our conclusions to fluences of 10% and higher. The edge length of the water box is 47 Å.

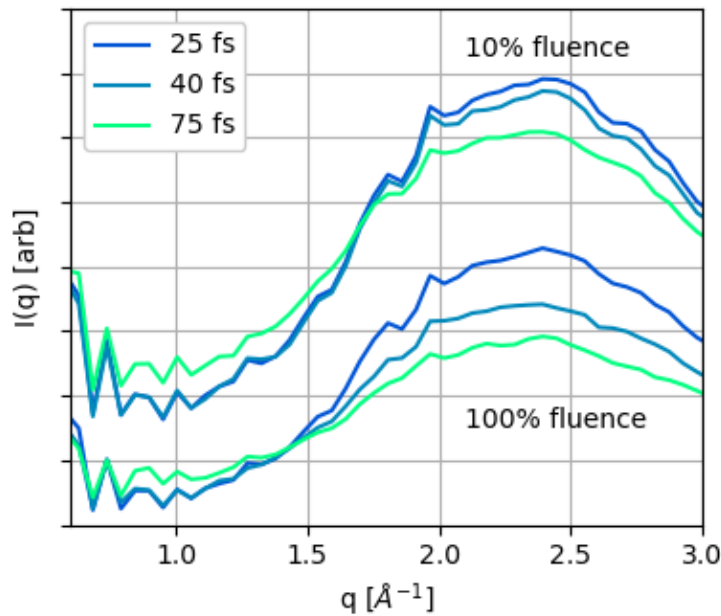

**Supplementary Figure S3. Simulation of the diffraction from single long pulses.** To facilitate comparison to the previously described experiment describing ultrafast water ionization[1], XMDYN simulations of single long pulses were performed and $I(q)$ traces estimated from the resulting time-integrated dynamics. Simulations of 25, 40, and 75 fs pulse duration were performed at 10% and 100% fluence, with fluence values based on the experiment presented here, not those from ref. [1]. In all cases, the water peak represents the predominant contribution to the scattering curves. For a 75 fs pulse there is some noticeable decrease of the ambient water peak and a slight rise in the low-$q$ region, especially at 10% fluence; the response, however, is significantly smaller than in the corresponding pump-probe simulations due to integration of the scattering signal across the entire pulse. This integration in time produces a signal with a large contribution from short delay times, where the atomic structure is still similar to that of the initial ambient water. The two sets of curves (100% and 10% fluence) are offset vertically for clarity.

**Supplementary Table 1: Fluence dependence of the calculated number of photoionization events in a water cube with 47 Å edge length triggered by the pump pulse.** The theoretical values were calculated for different nominal fluences using the photoionization cross-section and the total number of oxygen atoms in the box. 100 % fluence corresponds to $3.5 \times 10^{12}$ photons/µm² or an average intensity of $2.7 \times 10^{19}$ W/cm² ($5 \times 10^5$ J/cm²). To give a reference frame, the concentration of 31 ions in a water cube with 47 Å edge length corresponds to 0.5 M.

| Nominal fluence | Calculated number of photoionization events |
|---|---|
| 1 % | 4.82 |
| 1.5 % | 7.23 |
| 2 % | 9.64 |
| 2.5 % | 12.1 |
| 10 % | 48.2 |
| 50% | 241 |
| 100 % | 482 |

**Supplementary Table 2. Experimental parameters of selected XFEL XPCS experiments on water.**

| Photon Energy [keV] | Pulse Duration [fs] | Time Delay | Focus | Fluence per pulse | $q$-range | Comment | Publication |
|---|---|---|---|---|---|---|---|
| 8.2 | 10 to 120 | Not reported | 2 µm diameter | At sample: $10^9 - 10^{10}$ ph/pulse<br>Assuming Ø is FWHM, focus: $1.59 \times 10^8 - 1.59 \times 10^9$ ph/µm2<br>Peak central fluence $2.21 \times 10^8 - 2.21 \times 10^9$ ph/µm2 | 1.85-2.05 Å$^{-1}$ | Speckle Visibility Spectroscopy exposure time δt | Perakis et al., 2018[2] |
| 7.9 | Not reported | 1.3 ns | 16 µm diameter | 5 x$10^7$ photons/pulse at sample | 0.2-1.4 nm$^{-1}$ | | Roseker et al., 2018[3] |
| 9.3 | 50 | 886 ns | 4.4×3.6 µm$^2$ | 1.3- 56.8 mJ/mm$^2$<br>8.7x $10^5$- 3.81x$10^7$ ph/um$^2$ at sample | 0.1-0.6 nm$^{-1}$ | | Lehmkühler et al., 2020[4] |
| 10.0 | Not reported | 0-2 ps | 0.7×0.9 µm$^2$ | 4.86 x $10^9$ ph/um$^2$ (assuming all photons in the focus),<br>2.43 x $10^9$ ph/um$^2$ (half of the photons in the focus) | 1.0-4.0 Å$^{-1}$ | Focus on water peak (~ 2 Å$^{-1}$) Speckle Visibility Spectroscopy | Shinohara et al.,2020[5]; Zarkadoula et al., 2022[6] |
| 9.0 | ≤ 50 | 443 ns and 886 ns | 10 µm FWHM | 6.5 x $10^8$ ph/pulse at sample<br>4.14 x $10^6$ ph/µm$^2$ | 0.1-0.6 nm$^{-1}$ | MHz-XPCS on protein in aqueous solution | Reiser et al., 2022[7] |